\documentclass[
 reprint,
superscriptaddress,
 amsmath,amssymb,
 aps,
 pra,
longbibliography
]{revtex4-2}

\usepackage{ulem}

\usepackage{graphicx}
\usepackage{dcolumn}
\usepackage{bm}
\usepackage{xcolor}
\usepackage{physics}
\usepackage{hyperref}

\newcommand{\ee}{\mathrm{e}}

\begin{document}

\preprint{APS/123-QED}

\title{
Experimental subdiffraction source discrimination enabled by\\
spatial demultiplexing and single-photon detectors
}

\author{Luigi Santamaria Amato}
\affiliation{Agenzia Spaziale Italiana ASI, Contrada Terlecchia snc 75100, Matera, Italy
}

\author{Danilo Triggiani}
\email{danilo.triggiani1@gmail.com}
\affiliation{Dipartimento Interateneo di Fisica, Politecnico di Bari, 70126, Bari, Italy
}
\affiliation{Istituto Nazionale di Fisica Nucleare (INFN), Sezione di Bari, 70126 Bari, Italy}

\author{Cosmo Lupo}
\affiliation{Dipartimento Interateneo di Fisica, Politecnico di Bari, 70126, Bari, Italy
}
\affiliation{Istituto Nazionale di Fisica Nucleare (INFN), Sezione di Bari, 70126 Bari, Italy}

\date{\today}

\begin{abstract}
We experimentally demonstrate a universal, parameter-independent test for asymmetric source discrimination below the diffraction limit. The test exploits spatial-mode demultiplexing (SPADE) and single-photon detection to discriminate the presence of a faint source in the vicinity of a brighter one. We develop a complete theory of the test in the presence of arbitrary modal crosstalk and show that the crosstalk-affected test retains a favorable error exponent over diffraction-limited direct imaging, provided that the crosstalk remains below a threshold $C_{\mathrm{max}}\simeq 1/9$. This establishes a quantitative condition under which the subdiffraction advantage of SPADE survives realistic imperfections. We experimentally test the scheme for a range of source separations and intensity ratios, without requiring prior knowledge of these parameters. The measured false-negative error rates are in good agreement with the theoretical model and systematically outperform direct imaging. These results demonstrate that SPADE provides a robust and experimentally accessible advantage for asymmetric subdiffraction hypothesis testing in the photon-starved regime, even in the presence of realistic modal crosstalk.
\end{abstract}

\maketitle

Resolving images with high precision is a fundamental requirement across multiple fields and applications, such as bioimaging~\cite{Taylor2014, Schermelleh2019}, nanotechnologies~\cite{Pujals2019}, and astronomy~\cite{Kim2025}.
The natural limit on the resolution imposed in the optical domain by the diffraction of light passing through finite apertures has pushed the scientific community into a successful quest to find feasible `subdiffraction' imaging techniques~\cite{Hell1994,Gustafsson2000, Booth2026,Rust2006, Betzig2006, Napoli2019, Lelek2021}.
As an example, for exoplanet observations, resolving a faint planetary signal in close angular proximity to a bright host star poses a particularly stringent subdiffraction imaging problem~\cite{Guyon2018}.
In this context, established approaches such as coronagraphy~\cite{Mawet2017,Deshler2026} and starshades~\cite{Cash2006} aim to physically block the starlight to reduce the contrast ratio of the imaged scene.
In recent years, spatial-mode demultiplexing (SPADE)~\cite{Tsang2016, Lvovsky2026} has garnered significant attention as an alternative subdiffraction parameter estimation technique both theoretically~\cite{Lupo2016, Rehacek2017,Bisketzi2019, Oh2021} and experimentally~\cite{Zhou2019, Pushkina2021,Frank2023, Tan2023,  Amato23, Rouvire2024, Santamaria2025,  Wallis2025, Shringarpure2026} for its applicability to biosensing, surface metrology, and astronomical observations.
By projecting the incoming field onto an optimized set of transverse spatial modes and measuring them via photon counting, SPADE separates the signal and background at the level of the measurement, rather than through optical shading.
This approach exploits the efficiency of modern single-photon detection technologies, while being grounded on the theoretical framework of quantum information and sensing.

SPADE has been found to be a quantum-optimal technique for several tasks at the subdiffraction regime, including source discrimination~\cite{deAlmeida2021,Grace2022, Wadood2024,Triggiani2026} and in particular exoplanet detection~\cite{Lu2018, Huang2021}, where the goal is to correctly guess the presence of an exoplanet orbiting a star through statistical hypothesis testing (HT) on the observed data.
Employing the conventional terminology of standard binary decision HT, the absence or presence of an exoplanet constitutes the two hypotheses, $H_0$ and $H_1$ respectively.
Any test has a probability of error, either by guessing $H_1$ when $H_0$ is true (type I error or `false positive'), or by guessing $H_0$ when $H_1$ is true (type II error or `false negative'). 
Because of the rarity of exoplanets, a type II error is more costly than a type I.
Therefore, the problem of exoplanet detection is best studied within the framework of an asymmetric HT, that aims to minimize the type II error probability $\beta$ while keeping the type I error probability $\alpha$ bounded.
Within this framework, SPADE achieves a theoretical rate of false negatives well below what can be achieved by diffraction-limited direct imaging~\cite{Lu2018,Huang2021}.

In moving from theory to experiments, the presence of noise and modal crosstalk must be taken into consideration, since these drastically hinder the efficiency of SPADE-based measurements~\cite{Gessner2020, Amato23, deAlmeida2021}. Specifically, crosstalk is quantified by a nonvanishing probability of observing photons in a spatial mode at the output of the demultiplexer that is different from the input signal, and it is mainly caused by the imperfections in the device fabrication process.
Naive tests that work for ideal demultiplexers, generally based on checking whether photons can be detected in higher order modes, do not work in the presence of crosstalk, even if small.
In this direction, some authors have proposed tests that tolerate crosstalk.
Schlichtholz et al.~\cite{Schlichtholz2024} have proposed a crosstalk-tolerant test for sources of equal intensities that requires some prior knowledge on the source separation. 
Zanforlin et al.~\cite{Zanforlin2022} provided a proof-of-principle HT experiment using a two mode interferometer that is suboptimal with respect to SPADE and requires prior knowledge of the physical parameters.
Linowski et al.~\cite{Linowski2025} have theoretically proposed a test that is crosstalk-tolerant and does not depend on the intensity ratio between the two sources. 
Although SPADE-based HT employing a response thresholds calibrated on the measurement apparatus was recently performed~\cite{Wallis2025SuperresolvingBH}, an experimental demonstration of the advantage provided by SPADE in source-independent asymmetric HT in the presence of crosstalk is still lacking.

This work fills this gap.
The experiment is supported by a complete theory of HT including a generally noninvertible model of crosstalk.
The test we implement is an adaptation to our crosstalk model of the test theoretically proposed in Ref.~\cite{Linowski2025}, which is universal as it does not depend on the physical parameters.
We show that the test beats the subdiffraction limit, as it achieves a type II error probability $\beta$ that is lower than the error rate of an ideal direct-imaging-based test.
We find a `superresolution threshold' for the performance of SPADE-based detection below which SPADE outperforms direct imaging.
We show that the probability of incorrectly observing a photon in higher modes is the major cause of reduction of the error rate, which must be smaller than the threshold value $C_{\mathrm{max}}=1/9$ for SPADE to outperform direct imaging.  
As a comparison, we estimated the crosstalk of our demultiplexer $C\simeq 0.01$ to be one order of magnitude lower than this threshold, while state-of-the-art mode demultiplexers can achieve two orders of magnitude smaller crosstalk factors $C\simeq 3\times10^{-3}$~\cite{Santamaria22}.

\textsf{Theory and methods.}
Let us first review the theory of asymmetric HT applied to exoplanet detection as known in the literature and then introduce our model of crosstalk and theoretical results. Employing the formalism of quantum detection theory, the two hypotheses of absence $H_0$ or presence $H_1$ of the exoplanet are represented by two density matrices $\rho_0$ and $\rho_1$.
Let $\epsilon$ be the intensity ratio $I_B/I_A$ between the exoplanet and the main source, $d_a$ their distance normalized over the width of the point-spread function (PSF) of the optical system, which for simplicity we assume to be Gaussian, with $\epsilon,d_a\ll1$.
Let $\alpha$ and $\beta$ be the probabilities of type I and type II errors, respectively, of a given test.
In asymmetric HT, a quantum optimal test over an asymptotically large number $N\gg1$ of copies attains, by the quantum Stein lemma, a type II error probability $\beta\simeq\exp(-N D_\mathrm{Q}(\rho_0||\rho_1))$ that decays exponentially with $N$~\cite{Hiai1991, Ogawa2000}
where 
\begin{equation}
D_\mathrm{Q}(\rho_0||\rho_1)	\simeq \epsilon d_a^2/4
\label{eq:QRE}
\end{equation}
is the quantum relative entropy between $\rho_0$ and $\rho_1$, representing the error exponent of the test~\cite{Huang2021}.
In the same asymmetric HT scenario, any given test based on a measurement whose outcomes $x$ follow a probability distribution $p_i(x) \equiv p(x|H_i)$, conditioned on hypothesis $H_i$, attains for large $N$ at best $\beta\simeq \exp(-N D(p_0||p_1))$,
where $D(p_0||p_1)\leqslant D_\mathrm{Q}(\rho_0||\rho_1)$ is the relative entropy between the two distributions.

In the remainder of this paper, we identify with $N$ the total number of photons that can be detected in a given integration time. 
A test based on direct imaging with an ideal camera attains a relative entropy~\cite{Linowski2025}
\begin{equation}
	D_\mathrm{DI}(p_0||p_1)\simeq \epsilon^2d_a^4/4,
    \label{eq:DIRE}
\end{equation}
with a suboptimal scaling both in $d_a$ and $\epsilon$~\footnote{The hypotheses considered here are chosen so that they share the same centre of brightness.
This is compatible with a standard exoplanet detection scenario, in which the only prior knowledge on the system is the position of its centre~\cite{Linowski2025}.
In some other works, a different binary decision problem is formulated, in which the position of the star is known and common in the two hypotheses, while the centres of brightness are slightly offset~\cite{Huang2021}. 
This different choice of hypotheses leads to a different scaling in $d_a$ of the direct imaging relative entropy since the two hypotheses can be distinguished more easily via their offset centres. In Appendix~\ref{app:TwoBinaryTests} we discuss more in depth this discrepancy}.
On the other hand, ideal SPADE detection, projecting onto the Hermite-Gauss (HG) basis \{HG$_{ij}$\} for a Gaussian PSF, is known to saturate the quantum Stein lemma in Eq.~\eqref{eq:QRE}~\cite{Huang2021, Linowski2025}.
In particular, for small intensity ratios $\epsilon$ and normalized distances $d_a$, we can approximate the probabilities $p_j=(P(0|H_j),P(1|H_j))$ of observing under hypothesis $H_j$ a photon in mode $\text{HG}_{00}$ or a photon in either mode $\text{HG}_{01}$ or $\text{HG}_{10}$ as
\begin{equation}
    p_0=\begin{pmatrix}
        1\\
        0
    \end{pmatrix},\quad
    p_1=\begin{pmatrix}
        1-\epsilon d_a^2/4\\
         \epsilon d_a^2/4
    \end{pmatrix},
    \label{eq:ProbsSpade}
\end{equation} 
as shown in Appendix~\ref{app:SpadeProb}. Therefore, the outcomes follow Bernoulli distributions whose
relative entropy is quantum optimal, that is $D_\mathrm{SD}(p_0||p_1) = -\ln(P(0|H_1))\simeq \epsilon d_a^2/4 = D_\mathrm{Q}(\rho_0||\rho_1)$
\footnote{Notice that the two sources, if present, might not be aligned with the orientation of the demultiplexer, yielding non-vanishing and different probabilities of observing photons in modes HG$_{01}$ and HG$_{10}$, proportional to the projections of the relative position of the exoplanet onto the demultiplexer reference frame.
Measuring these probabilities separately is not required to reach quantum optimality. However, if the position of the exoplanet relative to the star is unknown, observing only one between the HG$_{01}$ and HG$_{10}$ would discard useful information and, in the worst case scenario, render the measurement completely blind to the presence of the exoplanet.}.

\begin{figure}[t]
    \centering
    \includegraphics[width=0.95\linewidth]{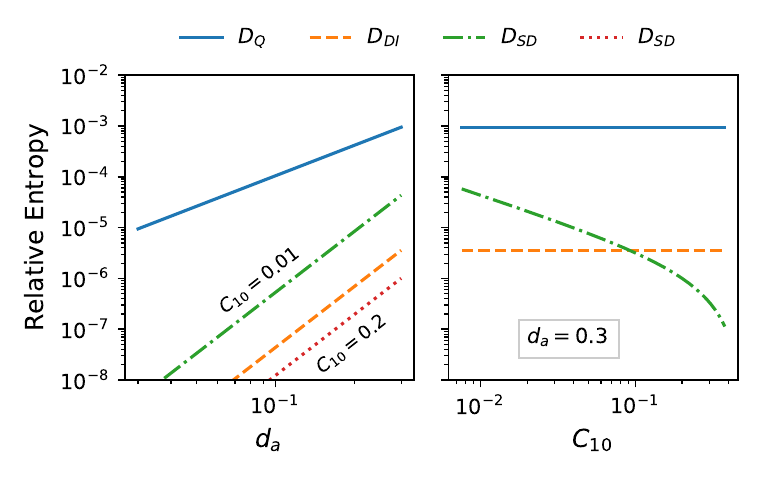} 
    \includegraphics[width=0.95\linewidth]{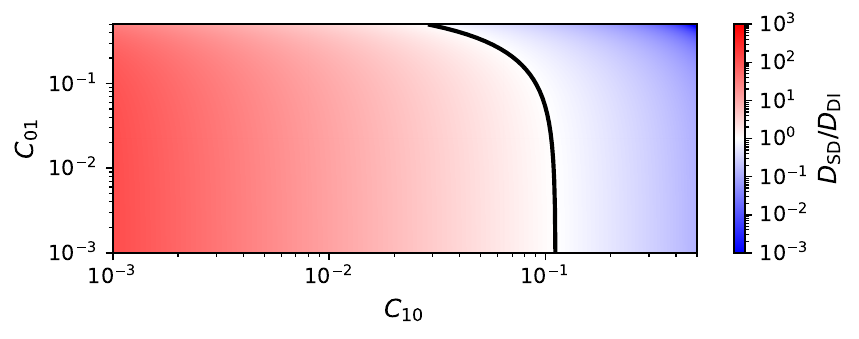}
    \caption{(Top) Plots of the quantum relative entropy $D_\mathrm{Q}$ in Eq.~\eqref{eq:QRE} (solid blue) and of the relative entropies for direct imaging $D_\mathrm{DI}$ in Eq.~\eqref{eq:DIRE} (orange dashed) and SPADE detection $D_\mathrm{SD}$ in Eq.~\eqref{eq:CrossRE} (green dash-dotted and red dotted) while varying the distances $d_a$ (left) and the crosstalk $C_{10}$ (right).
    $\epsilon=0.042$, $C_{01}=C_{10}$.
    (Bottom) Plot of the ratio $D_\mathrm{SD}/D_\mathrm{DI}$ while varying the crosstalk factors $C_{10}$ and $C_{01}$. The black curve delimits the region where SPADE outperforms direct imaging.}
    \label{fig:REs}
\end{figure}

We now introduce our  crosstalk model via an effective incoherent mixing of the outcome probabilities $p_j$ via a stochastic matrix,~\footnote{The effective crosstalk matrix generally depends on the physical crosstalk between the modes of the demultiplexer due to manufacturing imperfections and on the properties of the imaged system. We present in Appendix~\ref{app:Crosstalk} a model to retrieve the effective crosstalk from the instrument crosstalk.}
\begin{equation}
	p'_j=\begin{pmatrix}
	    P'(0|H_i) \\
        P'(1|H_i)
	\end{pmatrix} =
    \begin{pmatrix}
        C_{00} & C_{01} \\
        C_{10} & C_{11} 
    \end{pmatrix}
    \begin{pmatrix}
	    P(0|H_i) \\
        P(1|H_i)
	\end{pmatrix},
    \label{eq:ProbsCross}
\end{equation}
with $C_{0k}+C_{1k} = 1$ for $k=0,1$.
We can assume without loss of generality $C_{10},C_{01}\leq1/2$, where the equality corresponds to a purely noisy measurement.
Notice that generally, the transformation in Eq.~\eqref{eq:ProbsCross} is not invertible, and therefore extends to nonunitary channels, e.g.,~when the mode projection is coupled with some hidden and unobserved degrees of freedom.
We evaluate in Appendix~\ref{app:RE} the relative entropy between the two crosstalk-affected Bernoulli distributions with probabilities $p'_0$ and $p'_1$ for ${C_{10},C_{01}\gg\epsilon d_a^2}$,
\begin{equation}
	D_\mathrm{SD}(p'_0||p'_1) \simeq \frac{1}{32}\epsilon^2 d_a^4 \frac{(1-C_{01}-C_{10})^2}{C_{10}(1-C_{10})}\simeq\frac{\epsilon^2 d_a^4}{32 C_{10}},
    \label{eq:CrossRE}
\end{equation}
where in the last step we assume ${C_{10},C_{01}\ll 1}$~\cite{SM}.
The relative entropy $D_\mathrm{SD}(p'_0||p'_1)$ vanishes when $C_{10}=C_{11}$, i.e.~when the probability of observing a photon in either mode HG$_{10}$ or HG$_{01}$ becomes independent of the source.
By comparing $D_\mathrm{DI}$ in Eq.~\eqref{eq:DIRE} and $D_\mathrm{SD}$ in Eq.~\eqref{eq:CrossRE}, we see that SPADE asymptotically outperforms direct imaging in asymmetric HT, provided that the crosstalk satisfies
\begin{equation}
	\frac{(1-C_{01}-C_{10})^2}{32C_{10}(1-C_{10})}\geqslant\frac{1}{4}.
    \label{eq:CrosstalkInequality}
\end{equation}
In \figurename~\ref{fig:REs} we plot the relative entropies for SPADE and direct imaging, showcasing the higher performance of SPADE with respect to direct imaging in the low crosstalk regime.
The lower panel of \figurename~\ref{fig:REs} shows the existence of a `superresolution threshold' on the crosstalk below which SPADE outperforms direct imaging.
The factor $C_{10}$ plays a major role in determining the efficiency of SPADE, and there is a maximum value of crosstalk allowed $C_{10}\leqslant C_\mathrm{max}=1/9$, obtained imposing $C_{01}=0$ in Eq.~\eqref{eq:CrosstalkInequality}, for SPADE to outperform direct imaging, while any value of $C_{01}$ is acceptable.

We now describe the parameter-independent test based on SPADE measurement that achieves the optimal exponential scaling.
The test consists in comparing the number $N_1$ of photons observed overall in modes HG$_{01}$ and HG$_{10}$ with a threshold value $N_1^*$ fixed by the type I error probability $\alpha$ allowed, and the test rejects hypothesis $H_0$ if $N_1$ is larger than $N_1^*$.
We can find such a threshold 
as the value $N_1^*$ such that $P(N_1> N_1^*|H_0)=\alpha$, thereby fixing the type I error probability $\alpha$.
Notice that $N_1^*$ can either be measured in a previous calibration stage performed in absence of exoplanet, or exactly retrieved for large $N$ via the $\epsilon$- and $d_a$-independent expression
\begin{equation}
	N_1^*\simeq N C_{10}+K(\alpha) \sqrt{N C_{10} (1-C_{10})},
    \label{eq:Ntest}
\end{equation}
where $K(\alpha)=\sqrt{2}\ \mathrm{Erf}^{-1}(1-2\alpha)$, e.g. with $K(\alpha)=1.65$ for $\alpha=0.05$.
Indeed, one can easily check that the expression of $N_1^*$ in Eq.~\eqref{eq:Ntest} coincides with the $(1-\alpha)$th percentile of the large-$N$ Gaussian approximation of the binomial distribution $B(N,C_{10})$ associated with the photon count $N_1$ given hypothesis $H_0$.
We can calculate the type II error $\beta=P(N_1\leqslant N_1^*|H_1)$ in the presence of crosstalk for large $N$ if we approximate the binomial distribution $B(N,P'(1|H_1))$ given hypothesis $H_1$ to a Gaussian distribution with average $\mu_1=NP'(1|H_1)$ and variance $\sigma_1^2=NP'(1|H_1)P'(0|H_1)$.
For large $N$, we evaluate in Appendix~\ref{app:Calculations}
\begin{equation}
	\beta_{\mathrm{SD}} \simeq \frac{1}{2}\mathrm{Erfc}\left[\frac{\mu_1-N_1^*}{\sqrt{2}\sigma_1}\right]\simeq\ee^{-ND_\mathrm{SD}(p'_0||p'_1)},
    \label{eq:BetaTh}
\end{equation}
confirming the optimality of the test~\cite{SM}.

\begin{figure}
    \centering
	\includegraphics[width=1\linewidth]{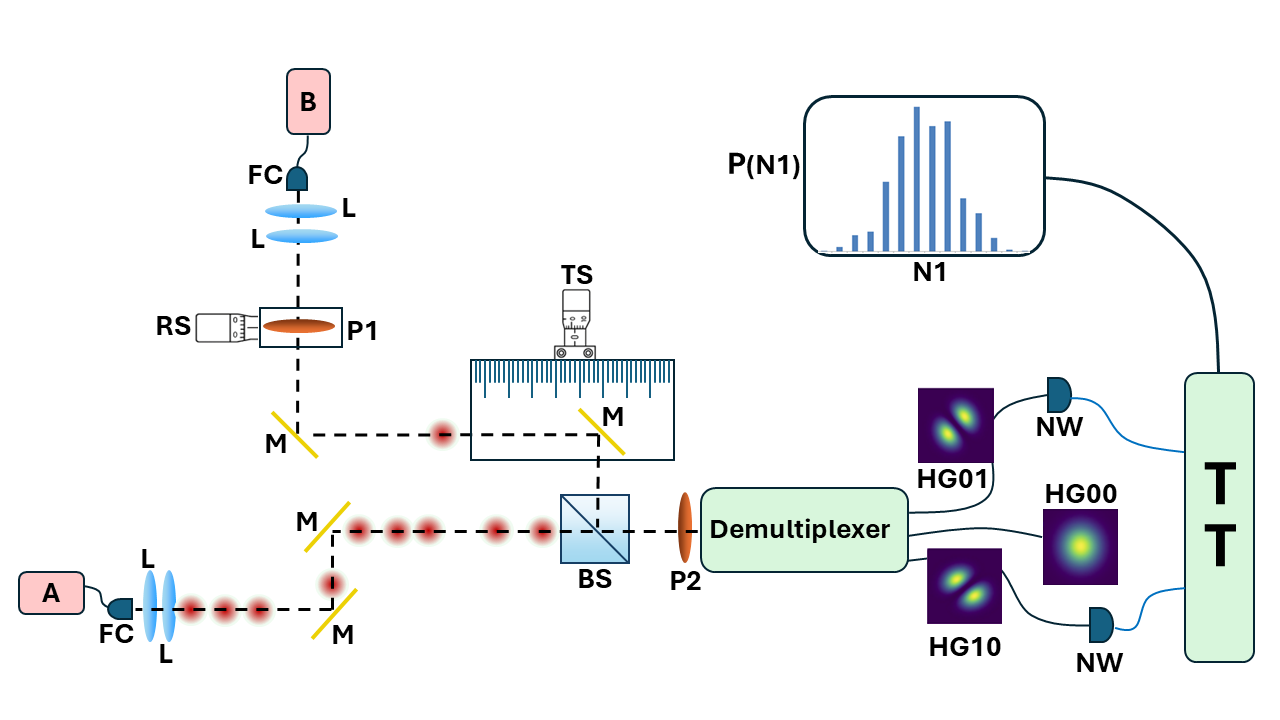}
	\caption{\label{setup} 
	Schematic representation of the measurement setup. See the main body of the paper for a detailed description.}
\end{figure}

\textsf{Experimental Setup.}
\figurename~\ref{setup} shows the experimental setup. Two fiber-coupled light emitting diodes (LEDs) with a central wavelength at about 1550 nm and a bandwidth of about 80 nm (FWHM, coherence time $\tau_c=\frac{0.44 \lambda^2}{c * FWHM}\sim 4 * 10 ^{-14} s $)
simulate two incoherent point-like sources. 
Both LEDs are free-space collimated using two fiber collimators (FCs) and a pair of two-lens (L) systems to match the beam waist $w_0$ of the demultiplexer adopted for the collection ($w_0 \simeq 300 \, \mu m$). The waist emulates the PSF of a hypothetical imaging system coupled with demultiplexer. 
LED A mimics the star and B the exoplanet; both are attenuated to, respectively, $I_A$ and $I_B$ intensities through fiber coupled attenuators not shown in the figure. 
The A beam, after reflection on two steering mirrors (M), impinges on a cube beam splitter (BS) to be combined with the B beam and, after crossing a fixed polarizer (P2), is coupled with the free-space input port of a HG demultiplexer (PROTEUS-C model from Cailabs with $300 \, \mu m$ input waist).
The B beam crosses a polarizer (P1)  mounted on a motorized rotation stage (RS) and two steering mirrors, the second of which is mounted on a micrometer translation stage (TS). 
It is then reflected on the BS, where it is combined with the A beam.
Both beams cross a fixed polarizer (P2) to overcome the dependency of detectors efficiency on photon polarization when the RS is rotated. 
The B beam is coupled with the HG demultiplexer too.
The translation stage moves the B beam and in turn changes the separation between the sources.
Furthermore, by rotating the RS, the  intensity of the B beam changes and in turn the intensity ratio $\epsilon = I_B/I_A$ changes.
The intensity and position of A stays fixed during the experiment.
Upon entering the demultiplexer, the two beams are first decomposed in the lowest-order HG modes:
$\text{HG}_{00}$, $\text{HG}_{01}$, $\text{HG}_{10}$,  
$\text{HG}_{11}$, $\text{HG}_{02}$, $\text{HG}_{20}$, and finally coupled with single-mode (SM) fibers.
The fibers corresponding to $\text{HG}_{01}$ and 
$\text{HG}_{10}$ are coupled to He-cooled superconducting nanowires (NWs) single-photon detectors  by means of  SM  fibers equipped with polarization paddles to align the photons’ polarization with the NW, whose detection efficiency is strongly polarization-dependent.
After the photons reach the nanowires, they generate electrical signals that, if exceeding the set threshold, are counted by a time tagger (TT) in the determined temporal window of $10$ ms.
The detector dark count is $20$ Hz, the system efficiency is about $80\%$ and reset time is about $100$ ns. 

\textsf{Results and discussion.} 
To measure the type II error probability $\beta$ of our setup, we repeat the test a number $M$ of times, and we estimate the real value of $\beta$ as the observed fail rate $\beta_{\mathrm{exp}}=(\text{number of fails})/M$, for different values of $\epsilon$, $d_a$ and $N$.
We choose $M=100$ in our analysis of $\beta_{\mathrm{exp}}$ while varying $d_a$, $M=200$ varying $\epsilon$ and $M=1000$ while varying $N$ to collect more statistics as we expect a smaller fail rate.
For each repetition,
we count the number $N_1$ of photons in the $\text{HG}_{01}$ and $\text{HG}_{10}$ modes in a fixed integration window of 10 ms. 
We calculate the thresholds $N_1^*$ required to perform the test as the $95$th percentile of the counts measured in absence of the exoplanet in a prior calibration stage, out of $M=300$ repetitions for the test with fixed $N$, and $M=1000$ for the test varying $N$, while we estimate its uncertainty via bootstrapping.
Notice that fixing the duration of the integration time, instead of the total number of counts, allows us to estimate $N_1^*$ employing only two detectors for the $\text{HG}_{10}$ and $\text{HG}_{01}$ modes, instead of the three normally required to also measure the counts in $\text{HG}_{00}$.
Although this experimental compromise introduces some statistical fluctuations in the total number of photons $N$ and generally reduces the performance of the test, we prove in Appendix~\ref{app:VacuumContr} that its effect is negligible in the small-crosstalk regime~\cite{SM}.

\begin{figure}
    \centering
    \includegraphics[width=0.95\linewidth]{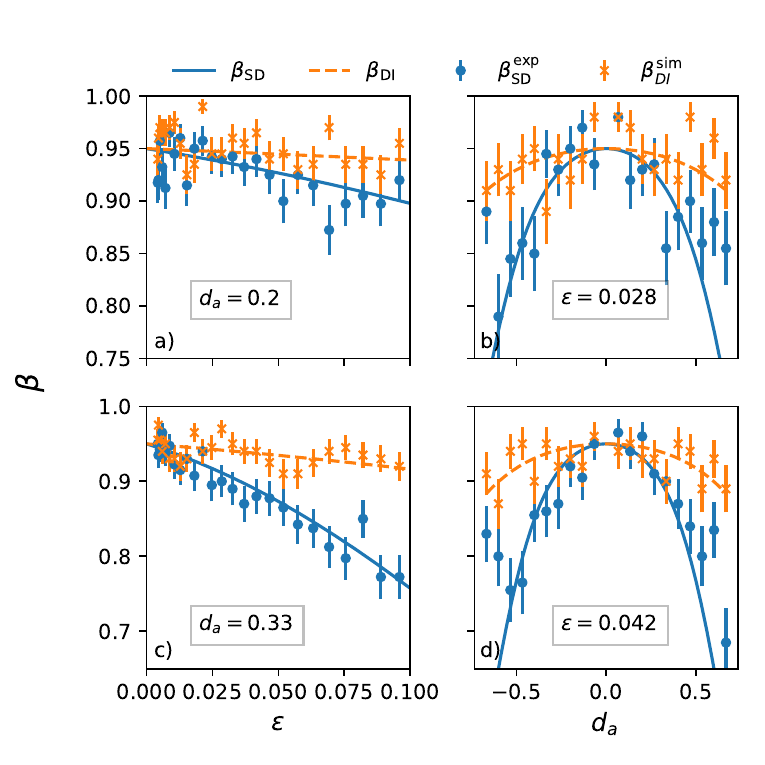}
    \includegraphics[width=0.95\linewidth]{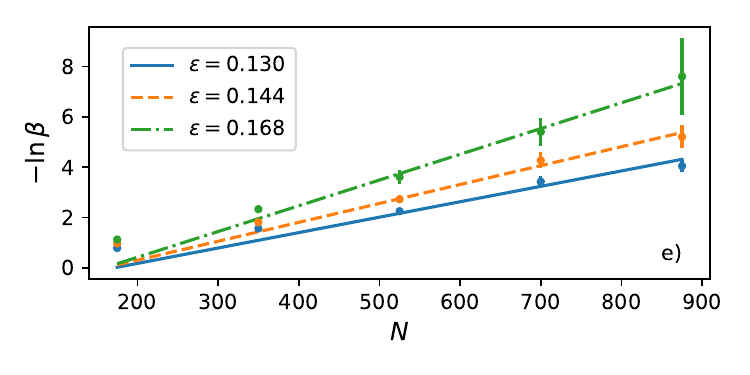}
    \caption{a)-d) Plots of the theoretically predicted error rate $\beta_\mathrm{SD}$ for SPADE in Eq.~\eqref{eq:BetaTh} (blue sold line) and $\beta_{DI}$ for direct imaging in Eq.~\eqref{eq:betaDI} (orange dashed line), of the error rates achieved with our experimental HT (blue dots), and of the error rates achieved by a Monte Carlo simulation of an ideal direct imaging test (orange crosses) for different values of the normalized distance $d_a$ and relative intensity $\epsilon$.
    e) Plots of the experimental error rates for increasing integration times with $d_a=0.67$.
    }
    \label{fig:REexp}
\end{figure}

The results are shown in \figurename~\ref{fig:REexp}. 
We compare the measured values of the error rates $\beta_\mathrm{exp}$ with the theoretical expression of $\beta_\mathrm{SD}$ in Eq.~\eqref{eq:BetaTh}, and with an ideal direct imaging test, described in S4 in the SM and simulated via Monte Carlo methods, yielding the error rate $\beta_\mathrm{DI}$~\cite{SM} \begin{equation}
	\beta_\mathrm{DI}=\frac{1}{2}\mathrm{Erfc}\left(\frac{\sqrt{N}\epsilon d_a^2-K(\alpha)\sqrt{2}}{2(1-\epsilon d_a^2)}\right)\simeq\ee^{-N D_\mathrm{DI}(p_0||p_1)}.
    \label{eq:betaDI}
\end{equation}
We observe a good match between theory and experiment, with the experimental SPADE-based test reliably outperforming the ideal direct-imaging-based simulated test. 
The uncertainties of the experimental points $\beta_{\mathrm{exp}}$ are estimated from the standard deviation of the binomial distribution $B(\beta_{\mathrm{exp}},M)$.
To obtain the theoretical curves for SPADE in panels a) to d) from Eq.~\eqref{eq:BetaTh}, we had to estimate $C_{10}$, $C_{01}$ and $N$.
To measure $N$, employing only the two available detectors, we coupled one with mode $\text{HG}_{00}$ and the other with mode $\text{HG}_{01}$ for $d_a=0$.
Then we counted the number of photons in the $\text{HG}_{00}$ mode plus twice the number of photons in the $\text{HG}_{10}$ mode in 10 ms as function of $\epsilon$ obtaining $N$ between $1010\pm34$ (when $\epsilon=0$) and $1095\pm37$ when ($\epsilon=0.096$). This procedure is justified by the fact that the demultiplexer was aligned at a 45° angle with the sources.
Known $N$, and evaluated $N_1^*=16\pm1$ from the data without exoplanet, we  could estimate $C_{10}\simeq0.010\pm0.001$ through Eq.~\eqref{eq:Ntest}.
Remarkably, this value of crosstalk falls within the region of $C_{10}<1/9$ for which SPADE detection largely outperforms direct imaging, as shown in Eq.~\eqref{eq:CrosstalkInequality} and \figurename~\ref{fig:REexp}. 
Finally, as expected from Eq.~\eqref{eq:CrossRE}, the dependency of the theoretical curve on $C_{01}$ is negligible for such small degrees of crosstalk, so we chose $C_{01}=C_{10}/2$ to be consistent with a balanced crosstalk effect, as per vendor specifications, according to our model of effective crosstalk described in S3 in the SM~\cite{SM}.
We notice from \figurename~\ref{fig:REexp} that the error rates predicted and measured are relatively high due to the short $10$ ms integration window, corresponding to $N\simeq10^3$ overall photon detected.
This is to be expected, since in our experimental conditions the scaling factor of the rate exponent is of the order of $\epsilon^2 d_a^4\lesssim10^{-5}$.

For the measurements for different integration times in \figurename~\ref{fig:REexp} e), we considered larger values of $\epsilon$ and $d_a$, with the scaling factor $\epsilon^2 d_a^4\simeq 4\times10^{-3}$, to achieve the asymptotic regime for smaller $N$.
Once again we first estimated the threshold values $N_1^*$, different for each $N$, from the counts in absence of exoplanet, obtaining the values $N_1^*=6,10,14, 17, 21$ and uncertainties estimated to be below the unit via bootstrapping.
In turn, from the experimental error rates $\beta_\mathrm{SD}^\mathrm{exp}$ for different values of $N$, we computed $C_{10}\simeq0.017\pm0.001$ by first determining $C_{10}$ from each of the linear regressions of the last three points and then averaging the resulting values. We then estimated $N$ via $N=N_1/C_{10}$ in the absence of an exoplanet, obtaining $N=848\pm34$ for the last data point. The curves in \figurename~\ref{fig:REexp}e) correspond to the linear regressions of the last three estimated $\beta_\mathrm{SD}^\mathrm{exp}$ points, with slope $D_\mathrm{SD}$ determined using the average value $C_{10}\simeq0.017$.
The larger value of crosstalk compared to the plots in \figurename~\ref{fig:REexp} a)-d), still comfortably lower than the superresolution threshold $C_{\mathrm{max}}\simeq 1/9$, can be caused by a different alignment of the setup.
Nevertheless, the expected exponential behaviour of $\beta$ is experimentally confirmed.

\textsf{Conclusions.}
We have theoretically analyzed and experimentally demonstrated a parameter-independent SPADE-based asymmetric HT for the detection of a secondary faint source in the subdiffraction regime. By assessing the measurement in terms of the relative-entropy error exponent, we have shown how modal crosstalk modifies the performance of the test and derived a general crosstalk model that also accounts for noninvertible, nonunitary transformations. The resulting test retains the exponential scaling of the false-negative error probability associated with the crosstalk-affected SPADE measurement, without requiring prior knowledge of the source separation or intensity ratio.
The analysis reveals a threshold of the crosstalk for SPADE to outperform direct imaging, and in particular a maximum value of crosstalk $C_{\mathrm{max}}=1/9$ allowed to reach a larger error exponent than direct imaging. In our experiment, the measured crosstalk is of order $10^{-2}$, well within this favorable regime.

These results establish that the subdiffraction advantage of SPADE for asymmetric HT is not restricted to ideal mode demultiplexers, but can persist under experimentally realistic crosstalk and imperfections. More broadly, our work provides the first experimental proof-of-concept of HG-SPADE-based asymmetric hypothesis testing, representing a key stepping stone toward applications on real observational data
\footnote{During the drafting of this work, we came across a more recent pre-print article performing an on-sky experimental realization of parameter-dependent asymmetric hypothesis testing using less precise double clad fiber as a binary mode sorter~\cite{Wallis2026}. Beside demonstrating the timeliness of our results, this work showcases the feasibility of on-sky SPADE measurements, and it is a further evidence that SPADE can achieve one or two orders-of-magnitude-increased performance compared with direct imaging, as we demonstrate in our experiment, if state-of-the-art detection setups and optimal hypothesis testing are employed with real observational data.}.
While atmospheric turbulence remains the principal challenge for ground-based astronomical observations and will require adaptive optics or space-based implementations, the demonstrated parameter-independent framework is directly applicable to a broad class of photon-limited subdiffraction detection problems. Beyond faint-source and exoplanet detection, an increasing number of applications of SPADE in fields outside astronomical observation are being recently proposed, such as microscopy~\cite{Kurdzialek2025} and nanoscale sensing for advanced manufacturing~\cite{Muller2026}, where reliable discrimination of weak or closely spaced optical signals is essential.

\textsf{Acknowledgments.} This work has received funding from the
European Union -- Next Generation EU, Missione 4 Componente 1, PRIN 2022, project title ``QUEXO'', CUP: D53D23002850006;
and the Italian Space Agency (Subdiffraction
Quantum Imaging ``SQI'' 2023-13-HH.0).

\textsf{Authors' contributions.} L.S.A. and D.T. contributed equally to this work. L.S.A. performed the experiment. D.T. developed the theoretical model and performed data analysis. L.S.A. and D.T. cowrote the manuscript. C.L. supervised the project and revised the manuscript.

\appendix

\section{Comparison between two exoplanet detection formulations}
\label{app:TwoBinaryTests}

In a standard exoplanet detection problem, one is interested in conceiving a test between two hypotheses: $H_0$ for the absence, $H_1$ for the presence of an exoplanet. 
However, there is an ambiguity in this formulation, that consists in what is known about the position of the two celestial bodies.
For example, Huang and Lupo~\cite{Huang2021} formulate the two hypotheses as
\begin{align*}
	H_0 &: \mathrm{One\ celestial\ body\ at\ position\ }x_s\\
    H_1 &: \mathrm{Star\ at\ position\ }x_s\ \mathrm{and\ planet\ at\ position\ } x_s+d,
\end{align*}
while Linowski et al.~\cite{Linowski2025} formulate the two hypotheses as
\begin{align*}
    H_0 : &\ \mathrm{One\ celestial\ body\ at\ position\ }x_0\\
    H_1 : &\ \mathrm{Star\ at\ position\ }x_s\ \mathrm{and\ planet\ at\ position\ } x_s+d,\\
    &\ \mathrm{but\ centre\ of\ brightness\ at\ }x_0.
\end{align*}
Although very similar, as they yield the same quantum relative entropy in the single-photon regime for small distances and intensity unbalances, these sets of hypotheses fundamentally differ in the prior knowledge on the system: in the first binary test, the position of the star is known, while in the second the position of the centre of brightness is known.
Here, we will evaluate the quantum relative entropy and the relative entropy associated with direct imaging of both binary tests, discussing the similarities and differences between the two.

Without loss of generality, we will consider a 1D system and Gaussian point-spread functions (PSFs).
The equal quantum relative entropy at the single-photon level (for small $\epsilon$ and $d_a$) in the two scenarios can be explained by the fact that it only depends, heuristically, on the overlap between the two states.
In the first binary decision, we can model the density matrices associated to the two hypotheses as
\begin{align}
    \rho_0 &= \ketbra{\psi(x_s)}{\psi(x_s)}\notag\\
    \rho_1 &= (1-\epsilon)\ketbra{\psi(x_s)}{\psi(x_s)}+\epsilon\ketbra{\psi(x_s+d_a)}{\psi(x_s+d_a)},
    \label{eq:RhoTest1}
\end{align}
for the second binary decision, as
\begin{align}
    \rho_0 &= \ketbra{\psi(x_0)}{\psi(x_0)}\notag\\
    \rho_1 &= (1-\epsilon)\ketbra{\psi(x_0-\epsilon d_a)}{\psi(x_0-\epsilon d_a)}+\notag\\
    &\quad+\epsilon\ketbra{\psi(x_0+(1-\epsilon)d_a)}{\psi(x_0+(1-\epsilon)d_a)},
    \label{eq:RhoTest2}
\end{align}
where $\ket{\psi(x)}$ represents a weak thermal source in position $x$ with PSF $\psi$,
so that the evaluation of  ${D_Q=-\mel{\psi(x_{s/0})}{\ln\rho_1}{\psi(x_{s/0})}}$ passes through evaluating $\ln\rho_1$, and therefore diagonalizing $\rho_1$.
For small $\epsilon$ and $d_a$, both versions of $\rho_1$ can be diagonalised in the Hermite-Gauss (HG) basis centred in the centre of brightness $x_0$.
In both cases we have
\begin{align}
	\rho_1 &= (1-v/4)\ketbra{0}{0} + v/4\ketbra{1}{1}\notag\\
    \ln\rho_1 &= \ln(1-v/4)\ketbra{0}{0} + \ln(v/4)\ketbra{1}{1}
\end{align}
where $v = \epsilon d_a^2$ is the variance of the intensity distribution of the two sources~\cite{Triggiani2026}, equal in the two $\rho_1$, while $\ket{i}$ is the $i$th HG mode centred in centre of brightness $x_0$.
We now notice that the star state $\rho_0$ coincides with the $\text{HG}_0$ mode in the second binary decision $\ket{\psi(x_0)}\equiv\ket{0}$, while $\ket{\psi(x_s)}\equiv\ket{0_{\epsilon d_a}}$, i.e.~the star state coincides with a $\text{HG}_0$ mode shifted of $-\epsilon d_a$.
For the second binary decision, the calculation is straightforward
\begin{equation}
	D_Q=-\mel{0}{\ln\rho_1}{0}\simeq \epsilon d_a^2/4,
\end{equation}
while for the first case
\begin{multline}
	D_Q =-\mel{0_{ed_a}}{\ln\rho_1}{0_{ed_a}}=\\ -\ln(1-v/4)|\braket{0}{0_{\epsilon d}}|^2 -\ln (v/4)|\braket{1}{0_{\epsilon d}}|^2\simeq \epsilon d_a^2/4
\end{multline}
since $\braket{0}{0_{\epsilon d}}\simeq1+\epsilon^2d_a^2$ and $\braket{1}{0_{\epsilon d}}\simeq \epsilon d$. 
The difference is negligible at the order $\epsilon d_a^2$.
However, notice that the terms we are neglecting are of order $\epsilon^2 d_a^2$, and not $\epsilon^2 d_a^4$.

Let us now consider a direct imaging setup.
The first binary test is compatible with the measured probability distributions proportional to the intensity profiles after the blur caused by the PSF (Gaussian with unitary variance)
\begin{align}
	p(x|H_0) &= \abs{\psi(x)}^2\notag\\
    p(x|H_1) &= (1-\epsilon)\abs{\psi(x)}^2 + \epsilon \abs{\psi(x-d_a)}^2.
\end{align}
Notice that the centre of brightness is different in the two hypotheses: $x_0\vert_{H_0}= 0$, $x_0\vert_{H_1} = \epsilon d_a$.
The relative entropy $D = \epsilon^2 d_a^2/2$ between these distributions scales (for small $\epsilon$ and $d_a$) quadratically in $d_a$. We can see that the scaling in $d_a$ is not the ``Rayleigh cursed" scaling. 
This is because the two hypothesis can be told apart by looking at the centre of brightness, although the relative entropy has suboptimal scaling in $\epsilon$.
The second binary test is compatible with probability distributions of the type
\begin{align}
	p(x|H_0) &= \abs{\psi(x)}^2\notag\\
    p(x|H_1) &= (1-\epsilon)\abs{\psi(x+\epsilon d_a)}^2 + \epsilon \abs{\psi(x-(1-\epsilon)d_a)}^2,
    \label{eq:ProbDI}
\end{align}
for which $x_0\vert_{H_0}=x_0\vert_{H_1}=0$.
The relative entropy $D = \epsilon^2 d_a^4/4$ has quartic scaling, that is, the ``Rayleigh cursed scaling'', and the quadratic scaling in $\epsilon$.

In Sec.~\ref{app:RE} we evaluate the relative entropies associated with SPADE measurement for both these binary test, and we show that, differently from direct imaging, the two binary tests are identical for small $\epsilon$ and $d_a$ under SPADE measurements.
In the next section, we will evaluate the probabilities associated to the outcomes of a SPADE measurement, required for the evaluation of the relative entropies.

\section{Spade probabilities}
\label{app:SpadeProb}

We recall that for a dim thermal source imaged with a Gaussian PSF in the single photon regime $\rho^{th}={(1-\eta)\ketbra{0}{0} +\eta\rho^{\{1\}}}$, the single-photon state can be written in the $2D$ HG basis
\begin{equation}
	\rho^{\{1\}}=\sum_{\substack{{n,n'}\\{m,m'}}=0}^\infty\gamma_{\substack{{n,n'}\\{m,m'}}}\ketbra{n,n'}{m,m'},
\end{equation}
with $\ket{n,n'}=\ket{n}_x\otimes\ket{n'}_y$ is the $(n,m)$-th HG mode, while
\begin{equation}
	\gamma_{\substack{{n,n'}\\{m,m'}}}=\int\dd x\dd y\ I(x,y)\frac{\ee^{-\frac{(x^2+y^2)}{4}}}{\sqrt{n!m!n'!m'!}}\left(\frac{x}{2}\right)^{n+m}\left(\frac{y}{2}\right)^{n'+m'},
    \label{eq:CovMat}
\end{equation}
where all spatial variables and the normalized distribution $I(x,y)$ at the source plane are expressed in units of the PSF width, considered unitary~\cite{Triggiani2026}.
Therefore, the probability of detecting the photon in the mode $\text{HG}_{j,k}$ is given by $P(j,k)=\gamma_{\substack{j,k\\j,k}}$.
For small intensity distributions (with respect to the PSF width), the only relevant modes are the lower ones, yielding
\begin{align}
	P(0,0)&\simeq\int\dd x\dd y\ I(x,y)(1-x^2/4-y^2/4)\notag\\
    &\qquad\qquad\qquad =1-M_{2x}-M_{2y}\\
    P(1,0)&\simeq\int\dd x\dd y\ I(x,y)(x^2/4)=M_{2x}\\
    P(0,1)&\simeq\int\dd x\dd y\ I(x,y)(y^2/4)=M_{2y},
\end{align}
where $M_{2x}$ and $M_{2y}$ are the second moments of the distribution $I(x,y)$.

For the exoplanet detection setup considered in the main text, we have two possible intensity distributions
\begin{align}
	I_0(x,y) &= \delta(x)\delta(y)\notag\\
    I_1(x,y) &= (1-\epsilon)\delta(x+\epsilon d_x)\delta(y+\epsilon d_y)\notag\\
    &\quad+\epsilon\delta(x-(1-\epsilon)d_x)\delta(y-(1-\epsilon)d_y)
\end{align}
associated with the two hypotheses of absence or presence of the exoplanet, where we fix the centre of brightness associated with $H_1$ and the star in $H_0$ in the origin of the reference frame, so that they are aligned with the demultiplexer, while the normalized distance between exoplanet and star is $d_a=\sqrt{d_x^2+d_y^2}$.
From Eq.~\eqref{eq:CovMat} it is clear that $H_0$ is associated with a pure state $\rho_0^{\{1\}}=\ketbra{0,0}{0,0}$, so that $P(0,0|H_0)=1$.
For hypothesis $H_1$, for small $d_a$ and $\epsilon$, we can easily evaluate $P(0,0|H_1)\simeq1-\epsilon d_a^2/4$, $P(1,0|H_1) \simeq \epsilon d_x^2/4$ and $P(0,1|H_1) \simeq \epsilon d_y^2/4$.
Therefore, if we consider the overall probability $P(1|H_1)=P(0,1|H_1)+P(1,0|H_1)$ to observe a photon in either mode $\text{HG}_{01}$ or $\text{HG}_{10}$, we obtain Eq.~\eqref{eq:ProbsSpade} of the main text.
Noticeably, we show in Sec.~\ref{app:RE} that the probabilities and relative entropies do not change if aligning the demultiplexer with the star, or if choosing the alternative binary test discussed in Sec.~\ref{app:TwoBinaryTests}.

\section{Crosstalk model}
\label{app:Crosstalk}

A physical model of crosstalk among all three relevant modes in our experiment is generally represented by a $3\times 3$ column-stochastic matrix $C_\mathrm{ph}$ so that
\begin{equation}
\begin{pmatrix}
    p'_1\\
    p'_2\\
    p'_3
\end{pmatrix}=
\begin{pmatrix}
c_{11} & c_{12} & c_{13}\\
c_{21} & c_{22} & c_{23}\\
c_{31} & c_{32} & c_{33}
\end{pmatrix}
\begin{pmatrix}
    p_1\\
    p_2\\
    p_3
\end{pmatrix}
\end{equation}
with $\sum_i(C_{\mathrm{ph}})_{ij}=1\ \forall j$, where we used a lightened notation to describe the three probabilities $(p_1,p_2,p_3)$ and $(p'_1,p'_2,p'_3)$ to observe a photon in modes (HG$_{00}$, HG$_{01}$, HG$_{10}$) in the ideal case and with crosstalk respectively.
However, we consider a test that counts the sum of the photons in modes HG$_{01}$ and HG$_{10}$.
We can therefore simplify the model, reducing the dimensionality of the stochastic matrix to two.
Let us consider a rectangular stochastic matrix 
\begin{equation}
	C_1=\begin{pmatrix}
	    1 & 0 & 0\\
        0 & 1& 1
	\end{pmatrix}
\end{equation}
that marginalizes over $p_2$ and $p_3$, so that $p'=C_1 C_\mathrm{ph}p$, i.e.
\begin{multline}
	\begin{pmatrix}
    p'_1\\
    p'_2+p'_3
\end{pmatrix}=
\begin{pmatrix}
	    1 & 0 & 0\\
        0 & 1& 1
	\end{pmatrix}
\begin{pmatrix}
c_{11} & c_{12} & c_{13}\\
c_{21} & c_{22} & c_{23}\\
c_{31} & c_{32} & c_{33}
\end{pmatrix}
\begin{pmatrix}
    p_1\\
    p_2\\
    p_3
\end{pmatrix}\\
=
\begin{pmatrix}
c_{11} & c_{12} & c_{13}\\
c_{21}+c_{31} & c_{22}+c_{32} & c_{23}+c_{33}
\end{pmatrix}
\begin{pmatrix}
    p_1\\
    p_2\\
    p_3
\end{pmatrix},
\end{multline}
And the rectangular stochastic matrix
\begin{equation}
	C_2=\begin{pmatrix}
    1 & 0\\
    0 & \frac{p_2}{p_2+p_3}\\
    0 & \frac{p_3}{p_2+p_3}
\end{pmatrix}
\end{equation}
so that
\begin{equation}
	\begin{pmatrix}
    p_1\\
    p_2\\
    p_3
\end{pmatrix}
=
\begin{pmatrix}
    1 & 0\\
    0 & \frac{p_2}{p_2+p_3}\\
    0 & \frac{p_3}{p_2+p_3}
\end{pmatrix}
\begin{pmatrix}
    p_1\\
    p_2+p_3
\end{pmatrix}.
\end{equation}
We can evaluate the effective crosstalk matrix
\begin{equation}
	C = C_1 C_\mathrm{ph} C_2 = 
    \begin{pmatrix}
        c_{11} & \frac{p_2c_{12}+p_3c_{13}}{p_2+p_3}\\
        c_{21}+c_{31} & \frac{p_2(c_{22}+c_{32})+p_3(c_{23}+c_{33})}{p_2+p_3}
    \end{pmatrix},
    \label{eq:EffCross}
\end{equation}
so that we can identify, from the crosstalk coefficients in the main body
\begin{equation}
	C_{10} = c_{21}+c_{12}, \qquad C_{01} = \frac{p_2c_{12}+p_3c_{13}}{p_2+p_3}.
\end{equation}
Notice that the transformation of the probabilities ${p'=Cp}$ through the application of the effective crosstalk matrix in Eq.~\eqref{eq:EffCross} is well defined also when hypothesis $H_0$ holds and $p_1$ is the only non-vanishing probability, since only the first column of $C$ is relevant in this case.
Moreover, the first column does not depend on the probabilities $p_i$, so that the effective crosstalk matrix $C$ is the same regardless of which hypothesis is correct.
In our experimental setup, the demultiplexer was rotated of 45° with respect to the plane of the two sources, so that $p_2=p_3$, and we assume $c_{12}=c_{21}$ and $c_{13}=c_{31}$ based on the demultiplexer manufacturer specifications, so that we retrieve the condition $C_{10}=2C_{01}$.
Interestingly, we can use this model to express, in terms of physical crosstalk matrix $C_\mathrm{ph}$, the condition $C_{10}=C_{11}$ for which the relative entropy $D_{SD}$ vanishes, namely
\begin{equation}
	C_{10}=C_{11} \iff c_{11} = \frac{p_2c_{12}+p_3c_{13}}{p_2+p_3}.
    \label{eq:WorstCaseCross}
\end{equation}
Expectedly, a uniform crosstalk matrix $c$ with elements $c_{ij}=1/3$ satisfies the equality in Eq.~\eqref{eq:WorstCaseCross}.

\section{Relative entropies}
\label{app:RE}

Here we will evaluate the relative entropies associated with the SPADE measurement, first in absence and then in presence of crosstalk.
We will consider two configurations for both binary tests discussed in Sec.~\ref{app:TwoBinaryTests}, with the demultiplexer aligned with the centre of brightness first, and then aligned with the star, for a total of four different scenarios.
We will show that these scenarios are all equivalent for SPADE within the first relevant order of the expansions in $\epsilon$ and $d_a$.
This analysis is important since it justifies the simplified experimental setup employed in this paper, where the demultiplexer was aligned for simplicity with the star and not with the centre of brightness, whereas the latter is the more practically relevant scenario for exoplanet detection.

We can evaluate all the probabilities in the remainder of this section with the formalism developed in Sec.~\ref{app:SpadeProb}.
We recall that the first binary decision is between the two hypotheses having the star fixed at a known point as represented in Eq.~\eqref{eq:RhoTest1}, while the second binary decision is between the two hypotheses having the centre of brightness fixed at a known point as represented in Eq.~\eqref{eq:RhoTest2}.
For the first binary decision, the alignment of the demultiplexer to the star means projecting over $\rho_0=\ketbra{0}{0}$, which is known to be a quantum optimal strategy.
Indeed, the probability of observing a photon in mode $\text{HG}_{0}$ and $\text{HG}_1$ are, for small $d_a$, 
\begin{align*}
    p(0|H_0)&=1,\quad p(1|H_0)=0\\
   p(0|H_1)&=1-M_2/4,\quad p(1|H_1)=M_2/4,
\end{align*}
where $M_2=\int\dd x\ I(x) x^2$ is the second moment of the source distribution (before PSF blurring) in the demultiplexer reference frame.
In the case at hand, $M_2=\int\dd x\ x^2((1-\epsilon)\delta(x)+\epsilon\delta(x-d_a))=\epsilon d_a^2$, and therefore $D=-\ln(1-M_2/4)\simeq \epsilon d_a^2/4$.
If we instead align the interferometer to the centre of brightness, we have
\begin{align*}
    p(0|H_0)&=1-M_2^{\{0\}}/4,\quad p(1|H_0)=M_2^{\{0\}}/4\\
   p(0|H_1)&=1-M_2^{\{1\}}/4,\quad p(1|H_1)=M_2^{\{1\}}/4,
\end{align*}
where now $M_2^{\{i\}}$ is the second moment (centred in the centre of intensity $\epsilon d_a$) of the $i$th hypothesis, namely $M_2^{\{0\}}=\int\dd x\ x^2\delta(x+\epsilon d_a)=\epsilon^2 d_a^2\simeq 0$ and $M_2^{\{1\}}=\int\dd x\ x^2((1-\epsilon)\delta(x+\epsilon d_a)+\epsilon\delta(x-d_a(1-\epsilon))=(1-\epsilon)\epsilon^2 d_a^2+\epsilon (1-\epsilon)^2d_a^2\simeq\epsilon d_a^2$, retrieving once again $D\simeq\epsilon d_a^2/4$.

We can do similar considerations for the second binary decision.
This time, centring the interferometer in the centre of brightness corresponds to projecting onto $\rho_0$
\begin{align*}
    p(0|H_0)&=1,\quad p(1|H_0)=0\\
   p(0|H_1)&=1-M_2/4,\quad p(1|H_1)=M_2/4,
\end{align*}
with $M_2= \int\dd x\ x^2((1-\epsilon)\delta(x+\epsilon d_a)+\epsilon\delta(x-(1-\epsilon)d_a))=(1-\epsilon)\epsilon^2d^2_a+\epsilon(1-\epsilon)^2d_a^2\simeq \epsilon d_a^2$, yielding $D=\epsilon d_a^2/4$.
Finally, centring on the star we have
\begin{align*}
    p(0|H_0)&=1-M_2^{\{0\}}/4,\quad p(1|H_0)=M_2^{\{0\}}/4\\
   p(0|H_1)&=1-M_2^{\{1\}}/4,\quad p(1|H_1)=M_2^{\{1\}}/4,
\end{align*}
with $M_2^{\{0\}}=\int \dd x\ x^2\delta(x-\epsilon d_a)=\epsilon^2 d_a^2\simeq 0$, and with $M_2^{\{1\}}=\int \dd x\ x^2((1-\epsilon)\delta(x)+\epsilon\delta(x-d_a))=\epsilon d_a^2$, so that once again $D=\epsilon d_a^2/4$.
We notice that, neglecting small quantities of orders of $\epsilon^2$ and $d_a^4$, the probabilities for each of the analysed cases reduce to the same expression, i.e.~Eq.~\eqref{eq:ProbsSpade} in the main text.

If we introduce crosstalk, we know the scaling of the relative entropy worsen to $\epsilon^2 d_a^4$.
This forces us to consider higher order moments in the probabilities evaluated above.
In particular, the generic probability before crosstalk effects will be of the form (after normalization)
\begin{align*}
    p(0) &= \gamma_{0,0} = 1-M_2/4+M_4/16\\
    p(1) &= \gamma_{1,1} = M_2/4-M_4/16,
\end{align*}
with $M_4$ forth order moment, while after the effect of crosstalk
\begin{align*}
    p'(0)&= C_{00}-(C_{00}-C_{01})(M_2/4-M_4/16) \\
    p'(1)&= C_{10}-(C_{10}-C_{11})(M_2/4-M_4/16)=1-p'(0).
\end{align*}
In this expression, we noted that $M_4$ is generally of order $d_a^4$ and $M_2$ of order $d_a^2$, therefore terms containing $M_4^2$, $M_4 M_2$ and $M_2^3$ are negligible.
The relative entropy reads
\begin{equation}
	D=p'_0\ln(\frac{p'_0}{p'_1})+(1-p'_0)\ln(\frac{1-p'_0}{1-p'_1}),
\end{equation}
which after some calculations can be shown to be independent of the forth order moment (up to the non-negligible orders) and reads
\begin{equation}
	D \simeq \frac{(1-C_{10}-C_{01})^2}{C_{10}(1-C_{10})}\frac{(M_2^{\{0\}}-M_2^{\{1\}})^2}{32},
\end{equation}
where $M_2^{\{i\}}$ is the second moment of the distribution of the $i$th hypothesis.
Let us go through each of the four cases examined before, for both binary tests and alignments of the interferometer.

For the first binary decision, when centring the demultiplexer on the star, we have $M_2^{\{0\}}=0$, $M_2^{\{1\}}=\epsilon d_a^2$, while if centring in the centre of brightness we have $M_2^{\{0\}}=\epsilon^2 d_a^2$, $M_2^{\{1\}}=\epsilon d_a^2-\epsilon^2 d_a^2$.
For the second binary decision, centring the interferometer on the centre of brightness yields $M_2^{\{0\}}=0$, $M_2^{\{1\}}=\epsilon d_a^2-\epsilon^2 d_a^2$, while centring on the star $M_2^{\{0\}}=\epsilon^2 d_a^2$, $M_2^{\{1\}}=\epsilon d_a^2$.
In any case $(M_2^{\{0\}}-M_2^{\{1\}})^2\simeq \epsilon^2d_a^4$ plus higher orders, yielding
\begin{equation}
	D = \frac{(1-C_{10}-C_{01})^2}{C_{10}(1-C_{10})}\frac{\epsilon^2 d_a^4}{32},
\end{equation}
as shown in Eq.~\eqref{eq:CrossRE} in the main text.

It is important to notice that the order expansion in $\epsilon$ and $d_a$ are performed assuming that all the other quantities in consideration are larger than what we are neglecting.
In particular, if the crosstalk is very weak, $C_{10}\simeq \epsilon d_a^2 $, then the expansions are not valid since the order of various terms changes, e.g. $\epsilon^2 d_a^4/C_{10}\simeq \epsilon d_a^2$.
However, as discussed in Ref.~\cite{Linowski2025}, typical values of crosstalk are several orders of magnitude greater than $\epsilon d_a^2$ in realistic exoplanet detection scenarios.

\section{Theoretical error rate of the test}
\label{app:Calculations}

Here we evaluate the performance of the test employed in our experiment.
Let us call $N_1$ the number of photons observed in either mode $\text{HG}_{01}$ or mode $\text{HG}_{10}$ after $N$ repetitions.
The defined hypotheses test guesses $H_0$ if $N_1<N_1^*$, $H_1$ if $N_1>N_1^*$, and it flips a coin if $N_1=N_1^*$, where $N_1^*$ is the $95$th percentile of the measured distribution of photons detected in either mode $\text{HG}_{01}$ or mode $\text{HG}_{10}$ when hypothesis $H_0$ is true.
By definition of $N_1^*$, $\alpha=P(H_1|H_0) = 5\%$.
To evaluate $\beta=P(H_0|H_1)$, we note again that the probability to observe $N_1$ given hypothesis $H_1$ is a binomial distribution with expected value $\mu_1 = P'(1|H_1) N$ and variance $\sigma_1^2 = P'(1|H_1)P'(0|H_1) N $, which can be approximated to a Gaussian distribution for large $N$, namely $\mathcal{N}(\mu_1,\sigma^2_1)$.
Therefore, since $\beta=P(N_1\leqslant N_1^*|H_1)$, we have
\begin{equation}
    \beta=\int_{-\infty}^{N_1^*}\dd x\ \frac{1}{\sqrt{2\pi\sigma_1^2}}\ee^{-\frac{(x-\mu_1)^2}{2\sigma_1^2}} = \frac{1}{2}\mathrm{Erfc}\left(\frac{\mu_1-N_1^*}{\sqrt{2}\sigma_1}\right),
\end{equation}
with
\begin{widetext}
\begin{align}
	\frac{\mu_1-N_1^*}{\sqrt{2}\sigma_1}=\frac{\sqrt{N}\epsilon d_a^2(1-C_{01}-C_{10})-1.65\sqrt{C_{10}(1-C_{10})}}{\sqrt{2(C_{10}+\epsilon d_a^2(1-C_{01}-C_{10}))(1-C_{10}-\epsilon d_a^2(1-C_{10}-C_{01})}},
    \label{eq:ArgErfc}
\end{align}
\end{widetext}
where we have employed the definition of $N_1^*$ as $95$th percentile in Eq.~\eqref{eq:Ntest} of the main text.
Since we are interested in the asymptotic regime of large $N$, we can neglect terms of order $\mathcal{O}(N^0)$, and then expanding in orders of $\epsilon d^2_a$, we have
\begin{equation}
	\frac{\mu_1-N_1^*}{\sqrt{2}\sigma_1}=\sqrt{N}\epsilon d_a^2 \frac{1-C_{01}-C_{10}}{\sqrt{2C_{10} (1-C_{10})}}+\mathcal{O}(\epsilon^2d_a^4).
\end{equation}
Since $\mathrm{Erfc}(k\sqrt{x})/2\simeq\ee^{-k^2x}$ for $k>0$ and $x\rightarrow\infty$, assuming $C_{11}>C_{10}$, we retrieve $\beta \simeq \exp(-N D_{SD}(p_0||p_1))$ as shown in Eq.~\eqref{eq:BetaTh} of the main text.

Without crosstalk, i.e.~for $C_{10}=C_{01}=0$, $C_{00}=C_{11}=1$, we have $N_1^*=0$, therefore the test guesses $H_0$ only if no photons are detected in either mode 
$\text{HG}_{01}$ or mode $\text{HG}_{10}$.
This is an outcome that largely deviates from $\mu_1 = N\epsilon d_a^2$ for large $N$, therefore the Gaussian distribution does not correctly approximate the probability distribution of $N_1$ under $H_1$, that is instead better approximated by a Poisson distribution.
However, we can exactly calculate $\beta$ as
\begin{equation}
	\beta = P(N_1=0|H_1) =  (1-\epsilon d_a^2)^{N}\simeq \ee^{-N \epsilon d_a^2}.
\end{equation}

With similar steps, we can evaluate the error rate $\beta_{DI}$ for a optimal direct imaging test with $\alpha=0.05$.
Indeed, assuming infinite resolution of the detectors, the outcome of each measurement will follow a probability density function, depending on the hypothesis
\begin{align}
	p(x|H_0) &= \abs{\psi(x)}^2\notag\\
    p(x|H_1) &= (1-\epsilon)\abs{\psi(x+\epsilon d_a)}^2 + \epsilon \abs{\psi(x-(1-\epsilon)d_a)}^2,
\end{align}
as written in Eq.~\eqref{eq:ProbDI}, with $\abs{\psi}^2$ normal distribution.
For small $d_a$, $P(x|H_1)$ can be written expanding in orders of $d_a$ and $\epsilon$,
\begin{equation}
	p(x|H_1) \simeq \left(1-\frac{\epsilon d_a^2}{2}\right)\abs{\psi(x)}^2 +\frac{\epsilon d_a^2}{2}x^2\abs{\psi(x)}^2.
\end{equation}
The two distribution have the same centre of brightness since $\int\dd x\ xp(x|H_0)=\int\dd x\ xp(x|H_1)=0$.
The easiest way the two distributions can be distinguished is by measuring their width, namely their variance
\begin{equation}
	\sigma_0^2=\int\dd x\ x^2 p(x|H_0)=1,\, \sigma_1^2=\int\dd x\ x^2 p(x|H_1)=1+\epsilon d_a^2,
\end{equation}
which can be estimated as mean square error $\bar{\sigma}^2=\sum x_i^2/N$, for large $N$, where $x_i$ is the $i$th outcome of the measurement following the distribution $p(x|H_{0/1})$.
However, $\bar{\sigma}$ has statistical fluctuations, and in particular
\begin{equation}
	v_j:=\mathrm{Var}_{H_j}[\bar{\sigma}^2]=\frac{1}{N^2}[3N\sigma^4_j+N(N-1)\sigma^4_j]-\sigma_j^4=\frac{2}{N}\sigma_j^4.
\end{equation}
From the central limit theorem, we can approximate the distribution of $\bar{\sigma}^2$ to Gaussian distributions.
We therefore consider the following test: if $\bar{\sigma}^2<\sigma^{*2}=\sigma_0^2+K(\alpha)\sqrt{v_0}=1+K(\alpha)\ \sqrt{2/N}$ the test guesses $H_0$, otherwise it guesses $H_1$, with
$K(\alpha)=\sqrt{2}\ \mathrm{Erf}^{-1}(1-2\alpha)\simeq1.65$ for $\alpha=0.05$.
Similarly to the SPADE test, this test by definition fixes $\alpha$ since it is a right-tailed $(1-\alpha)$th percentile test on the probability distribution of $H_0$.
We can calculate $\beta$ as
\begin{equation}
	\beta_{DI} = \int_{-\infty}^{\sigma^{*2}}\dd y\ \frac{1}{\sqrt{2\pi v_1}}\ee^{-\frac{(y-\sigma_1^2)^2}{2v_1}}=\frac{1}{2}\mathrm{Erfc}\left(\frac{\sigma_1^2-\sigma^{*2}}{\sqrt{2v_1}}\right)
\end{equation}
with
\begin{equation}
	\frac{\sigma_1^2-\sigma^{*2}}{\sqrt{2}v_1}=\frac{\sqrt{N}\epsilon d_a^2-K(\alpha)\sqrt{2}}{2(1-\epsilon d_a^2)}\simeq \sqrt{N}\frac{\epsilon d_a^2}{2},
\end{equation}
thus obtaining the expression in Eq.~\eqref{eq:betaDI} of the main text, and the error rate scaling $\beta_{DI}=\exp(-N\epsilon^2 d_a^4/4)$. 
This confirms the asymptotic optimality of this direct imaging test, in the sense that is saturates the error rate given by the relative entropy.

\section{Effect of a measurement at a fixed integration window} \label{app:VacuumContr}

If the photon counts are performed at a fixed time interval $\Delta t$, the total number $N$ of photons will show statistical fluctuations due to the thermal nature of the sources.
Here we will take into consideration these fluctuations and show that, for small crosstalk, they have negligible effect on the performance of the test.

To consider the fluctuations of the total number of photons, we must include the vacuum contributions of the thermal states
\begin{equation}
    \rho^{\mathrm{th}}_i = (1-\eta)\ketbra{0}{0} + \eta \rho^{\{1\}}_i+\mathcal{O}(\eta^2),
\end{equation}
where $\eta = I\tau$ is the probability that a photon is detected within a single temporal mode of duration $\tau$ of the thermal source of intensity $I$, while $\rho^{\{1\}}_i$ is the single-photon state associated with the hypothesis $H_i$.
For simplicity we will assume that $\eta$ is small enough so that at most one photon is detected per detection window of the detectors, that we can therefore identify with the duration $\tau$ of the temporal mode.
The outcome of a SPADE detection that measures the modes $\text{HG}_{00}$ and either mode
$\text{HG}_{10}$ or mode $\text{HG}_{01}$ after an integration over a time interval $\Delta t$ will follow a multinomial distribution with three possible events (absence of photons, photon in $\text{HG}_{00}$, photon in 
either mode $\text{HG}_{10}$ or mode $\text{HG}_{01}$) with respective probabilities, in absence of crosstalk
\begin{equation}
	P_v(\mathrm{vac}|H_i) = 1-\eta,\quad P_v(j|H_i) = \eta P(j|H_i),
\end{equation}
with $P(j|H_i)$ found in Eq.~\eqref{eq:ProbsSpade} of the main text for $j,i=0,1$.
The effect of crosstalk will be the mixing of the probabilities $P_v(j|H_i)$, obtaining the new probabilities
\begin{equation}
	P'_v(\mathrm{vac}|H_i) = 1-\eta,\quad P'_v(j|H_i) = \eta P'(j|H_i),
\end{equation}
with the same $P'(j|H_i)$ found in Eq.~\eqref{eq:ProbsCross} of the main text.

The quantity $\Delta t/\tau$ represents the number of repetitions of the measurement. 
If $\Delta t/\tau\gg1$, we can once again apply the central limit theorem and approximate the distribution of the counts in either mode 
$\text{HG}_{01}$ or mode $\text{HG}_{10}$ to a Gaussian distribution with average $\mu_i=I\Delta tP'(1|H_i)$, and variance $\sigma^2_i = I\Delta tP'(1|H_i)(1-P_v'(1|H_i))$.
The $95$th percentile value $N_1^*=\mu_0+1.65\ \sigma_0$ of the distribution, when $H_0$ is true, is given by
\begin{equation}
	N_1^* = I \Delta t C_{10} + 1.65\sqrt{I\Delta t C_{10}(1-I\tau C_{10})}.
\end{equation}
If we identify $I \Delta t = N$, we can rewrite
\begin{equation}
	N_1^* = N C_{10} + 1.65\sqrt{N C_{10}(1-C_{10}+C_{10}(1-I\tau))}.
\end{equation}
Notice the difference between this expression of $N_1^*$ and the one in Eq.~\eqref{eq:Ntest} of the main text, as the larger fluctuations due to the presence of the vacuum increase the value of $\sigma_0$ and therefore of $N_1^*$.
Repeating similar steps to the ones in Sec.~\ref{app:Calculations}, we obtain
\begin{equation}
	\beta = \frac{1}{2}\mathrm{Erfc}\left(\frac{\mu_1-N_1^*}{\sqrt{2}\sigma_1}\right),
\end{equation}
with
\begin{equation}
	\frac{\mu_1-N_1^*}{\sqrt{2}\sigma_1}\simeq\sqrt{N}\epsilon d_a^2 \frac{C_{11}-C_{10}}{\sqrt{2C_{10} (1-C_{10}+C_{10}(1-I\tau))}}.
\end{equation}
and therefore a relative entropy
\begin{equation}
	D_{SD}(p'_{v0}||p'_{v1})=\epsilon^2 d_a^4 \frac{(1-C_{01}-C_{10})^2}{32{C_{10} (1-C_{10}I\tau)}},
\end{equation}
which in general is smaller than $D_{SD}(p'_{0}||p'_{1})$ in Eq.~\eqref{eq:CrossRE}.
However, in the regime $\epsilon d_a^2\ll C_{10}\ll1$, we obtain the same exponent
\begin{equation}
	D_{SD}(p'_{v0}||p'_{v1})= \frac{\epsilon^2 d_a^4}{32C_{10} }.
\end{equation}

\bibliography{apssamp.bib}

@misc{SM,
  note = {See Supplemental Material at [URL will be inserted by publisher] for the discussion on the two different binary tests, derivations of the probability distributions associated with SPADE, our crosstalk model, relative entropies, theoretical error rates and the effect of measuring fixing the integration window instead of the overall photon count.}
}

@misc{Shringarpure2026,
      title={Experimental Evidence-Based Sub-Rayleigh Source Discrimination}, 
      author={Saurabh U. Shringarpure and Yong Siah Teo and Hyunseok Jeong and Michael Evans and Luis L. Sanchez-Soto and Antonin Grateau and Alexander Boeschoten and Nicolas Treps},
      year={2026},
      eprint={2601.13972},
      archivePrefix={arXiv},
      primaryClass={quant-ph},
      url={https://arxiv.org/abs/2601.13972}, 
}

@article{Wallis2025SuperresolvingBH,
  title={Super-resolving binary-source hypothesis testing with a double-clad fiber coupler.},
  author={John S Wallis and David R Gozzard and Alexander Frost and Benjamin P. Dix-Matthews and Nicolas G. J. Maron and Joshua J. Collier},
  journal={Optics letters},
  year={2025},
  volume={50 23},
  pages={
          7191-7194
        },
  url={https://api.semanticscholar.org/CorpusID:281951145}
}

@misc{Kurdzialek2025,
      title={Super-resolution microscopy via fluctuation-enhanced spatial mode demultiplexing}, 
      author={Stanislaw Kurdzialek},
      year={2025},
      eprint={2511.20790},
      archivePrefix={arXiv},
      primaryClass={physics.optics},
      url={https://arxiv.org/abs/2511.20790}, 
}

@misc{Muller2026,
      title={Looking down the rabbit hole: Towards quantum optimal estimation of surface roughness}, 
      author={Quentin Muller and Tommaso Tufarelli and Madalin Guta and Katherine Inzani and Samanta Piano and Gerardo Adesso},
      year={2026},
      eprint={2603.16418},
      archivePrefix={arXiv},
      primaryClass={quant-ph},
      url={https://arxiv.org/abs/2603.16418}, 
}

@misc{Wallis2026,
      title={On-sky binary source hypothesis testing beyond the diffraction limit using spatial mode demultiplexing based detection}, 
      author={John S. Wallis and Ayden S. McCann and Joshua J. Collier and Lilani D. Toms-Hardman and Alex M. Frost and Benjamin P. Dix-Matthews and David R. Gozzard},
      year={2026},
      eprint={2606.18025},
      archivePrefix={arXiv},
      primaryClass={astro-ph.IM},
      url={https://arxiv.org/abs/2606.18025}, 
}

@article{Deshler2026,
  title = {Quantum limits of exoplanet detection and localization},
  author = {Deshler, Nico and Haffert, Sebastiaan and Ashok, Amit},
  journal = {Phys. Rev. A},
  volume = {113},
  issue = {4},
  pages = {042406},
  numpages = {23},
  year = {2026},
  month = {Apr},
  publisher = {American Physical Society},
  doi = {10.1103/g6kr-mqdj},
  url = {https://link.aps.org/doi/10.1103/g6kr-mqdj}
}

@misc{Lvovsky2026,
      title={Passive optical superresolution at the quantum limit}, 
      author={A. I. Lvovsky and Michael R. Grace and Saikat Guha and Mankei Tsang and Gerardo Adesso and Nicolas Treps},
      year={2026},
      eprint={2605.10767},
      archivePrefix={arXiv},
      primaryClass={quant-ph},
      url={https://arxiv.org/abs/2605.10767}, 
}

@ARTICLE{Ogawa2000,
  author={Ogawa, T. and Nagaoka, H.},
  journal={IEEE Transactions on Information Theory}, 
  title={Strong converse and Stein's lemma in quantum hypothesis testing}, 
  year={2000},
  volume={46},
  number={7},
  pages={2428-2433},
  doi={10.1109/18.887855}}

@Article{Hiai1991,
author={Hiai, Fumio
and Petz, D{\'e}nes},
title={The proper formula for relative entropy and its asymptotics in quantum probability},
journal={Communications in Mathematical Physics},
year={1991},
month={Dec},
day={01},
volume={143},
number={1},
pages={99-114},
issn={1432-0916},
doi={10.1007/BF02100287},
url={https://doi.org/10.1007/BF02100287}
}

@article{Bisketzi2019,
doi = {10.1088/1367-2630/ab58a0},
url = {https://dx.doi.org/10.1088/1367-2630/ab58a0},
year = {2019},
month = {dec},
publisher = {IOP Publishing},
volume = {21},
number = {12},
pages = {123032},
author = {Bisketzi, Evangelia and Branford, Dominic and Datta, Animesh},
title = {Quantum limits of localisation microscopy},
journal = {New Journal of Physics}
}

@article{Mawet2017,
  author  = {Mawet, D. and Ruane, G. J. and Serabyn, E. and Echeverri, D. and Klimovich, N. and Randolph, M. and Fucik, J. and Wallace, J. K. and Wang, J. and Vasisht, G. and Dekany, R. and Mennesson, B. and Choquet, E. and Delorme, J.-R. and Serabyn, E.},
  title   = {Observing Exoplanets with High-Dispersion Coronagraphy. II. Demonstration of an Active Single-Mode Fiber Injection Unit},
  journal = {Astrophysical Journal},
  volume  = {838},
  pages   = {92},
  year    = {2017},
  doi     = {10.3847/1538-4357/aa647f}
}

@article{Guyon2018,
   author = "Guyon, Olivier",
   title = "Extreme Adaptive Optics", 
   journal= "Annual Review of Astronomy and Astrophysics",
   year = "2018",
   volume = "56",
   number = "Volume 56, 2018",
   pages = "315-355",
   doi = "https://doi.org/10.1146/annurev-astro-081817-052000",
   url = "https://www.annualreviews.org/content/journals/10.1146/annurev-astro-081817-052000",
   publisher = "Annual Reviews",
   issn = "1545-4282",
   type = "Journal Article",
  }

@Article{Cash2006,
author={Cash, Webster},
title={Detection of Earth-like planets around nearby stars using a petal-shaped occulter},
journal={Nature},
year={2006},
month={Jul},
day={01},
volume={442},
number={7098},
pages={51-53},
issn={1476-4687},
doi={10.1038/nature04930},
url={https://doi.org/10.1038/nature04930}
}

@article{Oh2021,
  title = {Quantum Limits of Superresolution in a Noisy Environment},
  author = {Oh, Changhun and Zhou, Sisi and Wong, Yat and Jiang, Liang},
  journal = {Phys. Rev. Lett.},
  volume = {126},
  issue = {12},
  pages = {120502},
  numpages = {7},
  year = {2021},
  month = {Mar},
  publisher = {American Physical Society},
  doi = {10.1103/PhysRevLett.126.120502},
  url = {https://link.aps.org/doi/10.1103/PhysRevLett.126.120502}
}

@article{Pushkina2021,
  title = {Superresolution Linear Optical Imaging in the Far Field},
  author = {Pushkina, A. A. and Maltese, G. and Costa-Filho, J. I. and Patel, P. and Lvovsky, A. I.},
  journal = {Phys. Rev. Lett.},
  volume = {127},
  issue = {25},
  pages = {253602},
  numpages = {6},
  year = {2021},
  month = {Dec},
  publisher = {American Physical Society},
  doi = {10.1103/PhysRevLett.127.253602},
  url = {https://link.aps.org/doi/10.1103/PhysRevLett.127.253602}
}

@article{Frank2023,
author = {Jernej Frank and Alexander Duplinskiy and Kaden Bearne and A. I. Lvovsky},
journal = {Optica},
number = {9},
pages = {1147--1152},
publisher = {Optica Publishing Group},
title = {Passive superresolution imaging of incoherent objects},
volume = {10},
month = {Sep},
year = {2023},
url = {https://opg.optica.org/optica/abstract.cfm?URI=optica-10-9-1147},
doi = {10.1364/OPTICA.493718},
}

@article{Rehacek2017,
author = {J. Rehacek and M. Pa\'{u}r and B. Stoklasa and Z. Hradil and L. L. S\'{a}nchez-Soto},
journal = {Opt. Lett.},
number = {2},
pages = {231--234},
publisher = {Optica Publishing Group},
title = {Optimal measurements for resolution beyond the Rayleigh limit},
volume = {42},
month = {Jan},
year = {2017},
url = {https://opg.optica.org/ol/abstract.cfm?URI=ol-42-2-231},
doi = {10.1364/OL.42.000231},
}

@article{Lupo2016,
  title = {Ultimate Precision Bound of Quantum and Subwavelength Imaging},
  author = {Lupo, Cosmo and Pirandola, Stefano},
  journal = {Phys. Rev. Lett.},
  volume = {117},
  issue = {19},
  pages = {190802},
  numpages = {5},
  year = {2016},
  month = {Nov},
  publisher = {American Physical Society},
  doi = {10.1103/PhysRevLett.117.190802},
  url = {https://link.aps.org/doi/10.1103/PhysRevLett.117.190802}
}

@Article{Pujals2019,
author={Pujals, Silvia
and Feiner-Gracia, Natalia
and Delcanale, Pietro
and Voets, Ilja
and Albertazzi, Lorenzo},
title={Super-resolution microscopy as a powerful tool to study complex synthetic materials},
journal={Nature Reviews Chemistry},
year={2019},
month={Feb},
day={01},
volume={3},
number={2},
pages={68-84},
issn={2397-3358},
doi={10.1038/s41570-018-0070-2},
url={https://doi.org/10.1038/s41570-018-0070-2}
}

@Article{Lelek2021,
author={Lelek, Micka{\"e}l
and Gyparaki, Melina T.
and Beliu, Gerti
and Schueder, Florian
and Griffi{\'e}, Juliette
and Manley, Suliana
and Jungmann, Ralf
and Sauer, Markus
and Lakadamyali, Melike
and Zimmer, Christophe},
title={Single-molecule localization microscopy},
journal={Nature Reviews Methods Primers},
year={2021},
month={Jun},
day={03},
volume={1},
number={1},
pages={39},
issn={2662-8449},
doi={10.1038/s43586-021-00038-x},
url={https://doi.org/10.1038/s43586-021-00038-x}
}

@Article{Schermelleh2019,
author={Schermelleh, Lothar
and Ferrand, Alexia
and Huser, Thomas
and Eggeling, Christian
and Sauer, Markus
and Biehlmaier, Oliver
and Drummen, Gregor P. C.},
title={Super-resolution microscopy demystified},
journal={Nature Cell Biology},
year={2019},
month={Jan},
day={01},
volume={21},
number={1},
pages={72-84},
issn={1476-4679},
doi={10.1038/s41556-018-0251-8},
url={https://doi.org/10.1038/s41556-018-0251-8}
}

@Article{Rust2006,
author={Rust, Michael J.
and Bates, Mark
and Zhuang, Xiaowei},
title={Sub-diffraction-limit imaging by stochastic optical reconstruction microscopy (STORM)},
journal={Nature Methods},
year={2006},
month={Oct},
day={01},
volume={3},
number={10},
pages={793-796},
issn={1548-7105},
doi={10.1038/nmeth929},
url={https://doi.org/10.1038/nmeth929}
}

@article{Betzig2006,
author = {Eric Betzig  and George H. Patterson  and Rachid Sougrat  and O. Wolf Lindwasser  and Scott Olenych  and Juan S. Bonifacino  and Michael W. Davidson  and Jennifer Lippincott-Schwartz  and Harald F. Hess },
title = {Imaging Intracellular Fluorescent Proteins at Nanometer Resolution},
journal = {Science},
volume = {313},
number = {5793},
pages = {1642-1645},
year = {2006},
doi = {10.1126/science.1127344},
URL = {https://www.science.org/doi/abs/10.1126/science.1127344},
eprint = {https://www.science.org/doi/pdf/10.1126/science.1127344}}

@article{Gustafsson2000,
author = {Gustafsson, M. G. L.},
title = {Surpassing the lateral resolution limit by a factor of two using structured illumination microscopy},
journal = {Journal of Microscopy},
volume = {198},
number = {2},
pages = {82-87},
doi = {https://doi.org/10.1046/j.1365-2818.2000.00710.x},
url = {https://onlinelibrary.wiley.com/doi/abs/10.1046/j.1365-2818.2000.00710.x},
eprint = {https://onlinelibrary.wiley.com/doi/pdf/10.1046/j.1365-2818.2000.00710.x},
year = {2000}
}

@article{Hell1994,
author = {Stefan W. Hell and Jan Wichmann},
journal = {Opt. Lett.},
number = {11},
pages = {780--782},
publisher = {Optica Publishing Group},
title = {Breaking the diffraction resolution limit by stimulated emission: stimulated-emission-depletion fluorescence microscopy},
volume = {19},
month = {Jun},
year = {1994},
url = {https://opg.optica.org/ol/abstract.cfm?URI=ol-19-11-780},
doi = {10.1364/OL.19.000780},
}

@article{Napoli2019,
  title = {Towards Superresolution Surface Metrology: Quantum Estimation of Angular and Axial Separations},
  author = {Napoli, Carmine and Piano, Samanta and Leach, Richard and Adesso, Gerardo and Tufarelli, Tommaso},
  journal = {Phys. Rev. Lett.},
  volume = {122},
  issue = {14},
  pages = {140505},
  numpages = {7},
  year = {2019},
  month = {Apr},
  publisher = {American Physical Society},
  doi = {10.1103/PhysRevLett.122.140505},
  url = {https://link.aps.org/doi/10.1103/PhysRevLett.122.140505}
}

@article{Kim2025,
doi = {10.3847/2041-8213/ae0739},
url = {https://doi.org/10.3847/2041-8213/ae0739},
year = {2025},
month = {oct},
publisher = {The American Astronomical Society},
volume = {993},
number = {1},
pages = {L3},
author = {Kim, Yoo Jung and Fitzgerald, Michael P. and Vievard, Sébastien and Lin, Jonathan and Xin, Yinzi and Lucas, Miles and Guyon, Olivier and Lozi, Julien and Deo, Vincent and Huby, Elsa and Lacour, Sylvestre and Lallement, Manon and Amezcua-Correa, Rodrigo and Leon-Saval, Sergio and Norris, Barnaby and Nowak, Mathias and Sallum, Steph and Sarrazin, Jehanne and Taras, Adam and Yerolatsitis, Stephanos and Jovanovic, Nemanja},
title = {On-sky Demonstration of Subdiffraction-limited Astronomical Measurement Using a Photonic Lantern},
journal = {The Astrophysical Journal Letters}
}

@misc{Booth2026,
      title={Structured detection microscopy}, 
      author={Larnii Booth and Kyle Clunies-Ross and Rumelo Amor and Nicolas Mauranyapin and Zixin Huang and Michael A. Taylor and Warwick P. Bowen},
      year={2026},
      eprint={2604.00413},
      archivePrefix={arXiv},
      primaryClass={physics.optics},
      url={https://arxiv.org/abs/2604.00413}, 
}

@article{Taylor2014,
  title = {Subdiffraction-Limited Quantum Imaging within a Living Cell},
  author = {Taylor, Michael A. and Janousek, Jiri and Daria, Vincent and Knittel, Joachim and Hage, Boris and Bachor, Hans-A. and Bowen, Warwick P.},
  journal = {Phys. Rev. X},
  volume = {4},
  issue = {1},
  pages = {011017},
  numpages = {7},
  year = {2014},
  month = {Feb},
  publisher = {American Physical Society},
  doi = {10.1103/PhysRevX.4.011017},
  url = {https://link.aps.org/doi/10.1103/PhysRevX.4.011017}
}

@PREAMBLE{
 "\providecommand{\noopsort}[1]{}" 
 # "\providecommand{\singleletter}[1]{#1}%" 
}

@article{Huang2021,
  title = {Quantum Hypothesis Testing for Exoplanet Detection},
  volume = {127},
  ISSN = {1079-7114},
  url = {http://dx.doi.org/10.1103/PhysRevLett.127.130502},
  DOI = {10.1103/physrevlett.127.130502},
  number = {13},
  journal = {Physical Review Letters},
  publisher = {American Physical Society (APS)},
  author = {Huang,  Zixin and Lupo,  Cosmo},
  year = {2021},
  month = sep 
}

@article{Tsang2016,
  title = {Quantum Theory of Superresolution for Two Incoherent Optical Point Sources},
  volume = {6},
  ISSN = {2160-3308},
  url = {http://dx.doi.org/10.1103/PhysRevX.6.031033},
  DOI = {10.1103/physrevx.6.031033},
  number = {3},
  journal = {Physical Review X},
  publisher = {American Physical Society (APS)},
  author = {Tsang,  Mankei and Nair,  Ranjith and Lu,  Xiao-Ming},
  year = {2016},
  month = aug 
}

@article{Gessner2020,
  doi = {10.1103/physrevlett.125.100501},
  url = {https://doi.org/10.1103/physrevlett.125.100501},
  year = {2020},
  month = aug,
  publisher = {American Physical Society ({APS})},
  volume = {125},
  number = {10},
  pages = {100501},
  author = {Manuel Gessner and Claude Fabre and Nicolas Treps},
  title = {Superresolution Limits from Measurement Crosstalk},
  journal = {Physical Review Letters}
}

@article{Schlichtholz2024,
  title = {Practical tests for sub-Rayleigh source discriminations with imperfect demultiplexers},
  volume = {2},
  ISSN = {2837-6714},
  url = {http://dx.doi.org/10.1364/OPTICAQ.502459},
  DOI = {10.1364/opticaq.502459},
  number = {1},
  journal = {Optica Quantum},
  publisher = {Optica Publishing Group},
  author = {Schlichtholz,  Konrad and Linowski,  Tomasz and Walschaers,  Mattia and Treps,  Nicolas and Rudnicki,  Lukasz and Sorelli,  Giacomo},
  year = {2024},
  month = jan,
  pages = {29}
}

@article{Linowski2025,
  title = {Quantum-inspired exoplanet detection in the presence of experimental imperfections},
  volume = {24},
  ISSN = {2331-7019},
  url = {http://dx.doi.org/10.1103/qq99-jmpv},
  DOI = {10.1103/qq99-jmpv},
  number = {4},
  journal = {Physical Review Applied},
  publisher = {American Physical Society (APS)},
  author = {Linowski,  Tomasz and Schlichtholz,  Konrad and Sorelli,  Giacomo},
  year = {2025},
  month = oct 
}

@article{Santamaria22,
author = {Luigi Santamaria and Deborah Pallotti and Mario Siciliani de Cumis and Daniele Dequal and Cosmo Lupo},
journal = {Opt. Express},
number = {21},
pages = {33930--33944},
publisher = {Optica Publishing Group},
title = {Spatial-mode demultiplexing for enhanced intensity and distance measurement},
volume = {31},
month = {Oct},
year = {2023},
url = {https://opg.optica.org/oe/abstract.cfm?URI=oe-31-21-33930},
doi = {10.1364/OE.486617},
}

@article{Zanforlin2022,
  title = {Optical quantum super-resolution imaging and hypothesis testing},
  volume = {13},
  ISSN = {2041-1723},
  url = {http://dx.doi.org/10.1038/s41467-022-32977-8},
  DOI = {10.1038/s41467-022-32977-8},
  number = {1},
  journal = {Nature Communications},
  publisher = {Springer Science and Business Media LLC},
  author = {Zanforlin,  Ugo and Lupo,  Cosmo and Connolly,  Peter W. R. and Kok,  Pieter and Buller,  Gerald S. and Huang,  Zixin},
  year = {2022},
  month = sep 
}

@article{Wadood2024,
  title = {Experimental demonstration of quantum-inspired optical symmetric hypothesis testing},
  volume = {49},
  ISSN = {1539-4794},
  url = {http://dx.doi.org/10.1364/OL.512320},
  DOI = {10.1364/ol.512320},
  number = {3},
  journal = {Optics Letters},
  publisher = {Optica Publishing Group},
  author = {Wadood,  S. A. and Sethuraj,  K. R. and Liang,  Kevin and Grace,  Michael R. and La Rue,  Gavin and Guha,  Saikat and Vamivakas,  A. N.},
  year = {2024},
  month = jan,
  pages = {750}
}

@article{Lu2018,
  title = {Quantum-optimal detection of one-versus-two incoherent optical sources with arbitrary separation},
  volume = {4},
  ISSN = {2056-6387},
  url = {http://dx.doi.org/10.1038/s41534-018-0114-y},
  DOI = {10.1038/s41534-018-0114-y},
  number = {1},
  journal = {npj Quantum Information},
  publisher = {Springer Science and Business Media LLC},
  author = {Lu,  Xiao-Ming and Krovi,  Hari and Nair,  Ranjith and Guha,  Saikat and Shapiro,  Jeffrey H.},
  year = {2018},
  month = dec 
}

@article{deAlmeida2021,
  title = {Discrimination and estimation of incoherent sources under misalignment},
  volume = {103},
  ISSN = {2469-9934},
  url = {http://dx.doi.org/10.1103/PhysRevA.103.022406},
  DOI = {10.1103/physreva.103.022406},
  number = {2},
  journal = {Physical Review A},
  publisher = {American Physical Society (APS)},
  author = {de Almeida,  J. O. and Kołodyński,  J. and Hirche,  C. and Lewenstein,  M. and Skotiniotis,  M.},
  year = {2021},
  month = feb 
}

@article{Grace2022,
  title = {Identifying Objects at the Quantum Limit for Superresolution Imaging},
  volume = {129},
  ISSN = {1079-7114},
  url = {http://dx.doi.org/10.1103/PhysRevLett.129.180502},
  DOI = {10.1103/physrevlett.129.180502},
  number = {18},
  journal = {Physical Review Letters},
  publisher = {American Physical Society (APS)},
  author = {Grace,  Michael R. and Guha,  Saikat},
  year = {2022},
  month = oct 
}

@article{Wallis2025,
  title = {Spatial mode demultiplexing for super-resolved source parameter estimation},
  volume = {33},
  ISSN = {1094-4087},
  url = {http://dx.doi.org/10.1364/OE.563503},
  DOI = {10.1364/oe.563503},
  number = {16},
  journal = {Optics Express},
  publisher = {Optica Publishing Group},
  author = {Wallis,  John S. and Gozzard,  David R. and Frost,  Alex M. and Collier,  Joshua J. and Maron,  Nicolas and Dix-Matthews,  Benjamin P.},
  year = {2025},
  month = aug,
  pages = {34651}
}

@article{Zhou2019,
  title = {Quantum-limited estimation of the axial separation of two incoherent point sources},
  volume = {6},
  ISSN = {2334-2536},
  url = {http://dx.doi.org/10.1364/OPTICA.6.000534},
  DOI = {10.1364/optica.6.000534},
  number = {5},
  journal = {Optica},
  publisher = {Optica Publishing Group},
  author = {Zhou,  Yiyu and Yang,  Jing and Hassett,  Jeremy D. and Rafsanjani,  Seyed Mohammad Hashemi and Mirhosseini,  Mohammad and Vamivakas,  A. Nick and Jordan,  Andrew N. and Shi,  Zhimin and Boyd,  Robert W.},
  year = {2019},
  month = apr,
  pages = {534}
}

@article{Tan2023,
  title = {Quantum-inspired superresolution for incoherent imaging},
  volume = {10},
  ISSN = {2334-2536},
  url = {http://dx.doi.org/10.1364/OPTICA.493227},
  DOI = {10.1364/optica.493227},
  number = {9},
  journal = {Optica},
  publisher = {Optica Publishing Group},
  author = {Tan,  Xiao-Jie and Qi,  Luo and Chen,  Lianwei and Danner,  Aaron J. and Kanchanawong,  Pakorn and Tsang,  Mankei},
  year = {2023},
  month = sep,
  pages = {1189}
}

@article{Amato23,
author = {Luigi Santamaria and Fabrizio Sgobba and Cosmo Lupo},
journal = {Optica Quantum},
number = {1},
pages = {46--56},
publisher = {Optica Publishing Group},
title = {Single-photon sub-Rayleigh precision measurements of a pair of incoherent sources of unequal intensity},
volume = {2},
month = {Feb},
year = {2024},
url = {https://opg.optica.org/opticaq/abstract.cfm?URI=opticaq-2-1-46},
doi = {10.1364/OPTICAQ.505457},
}

@article{Santamaria2025,
  title = {Single-photon super-resolved spectroscopy from spatial-mode demultiplexing},
  volume = {13},
  ISSN = {2327-9125},
  url = {http://dx.doi.org/10.1364/PRJ.544197},
  DOI = {10.1364/prj.544197},
  number = {4},
  journal = {Photonics Research},
  publisher = {Optica Publishing Group},
  author = {Santamaria,  Luigi and Sgobba,  Fabrizio and Pallotti,  Deborah and Lupo,  Cosmo},
  year = {2025},
  month = mar,
  pages = {865}
}

@article{Rouvire2024,
  title = {Ultra-sensitive separation estimation of optical sources},
  volume = {11},
  ISSN = {2334-2536},
  url = {http://dx.doi.org/10.1364/OPTICA.500039},
  DOI = {10.1364/optica.500039},
  number = {2},
  journal = {Optica},
  publisher = {Optica Publishing Group},
  author = {Rouviere,  Clementine and Barral,  David and Grateau,  Antonin and Karuseichyk,  Ilya and Sorelli,  Giacomo and Walschaers,  Mattia and Treps,  Nicolas},
  year = {2024},
  month = jan,
  pages = {166}
}

@article{Triggiani2026,
doi = {10.1088/2058-9565/ae2885},
url = {https://doi.org/10.1088/2058-9565/ae2885},
year = {2025},
month = {dec},
publisher = {IOP Publishing},
volume = {11},
number = {1},
pages = {015028},
author = {Triggiani, Danilo and Lupo, Cosmo},
title = {Achieving quantum-limited sub-Rayleigh identification of incoherent optical sources with arbitrary intensities},
journal = {Quantum Science and Technology}
}

\end{document}